\documentclass[9pt,twocolumn,twoside]{osajnl}
\journal{preprint}

\newcommand{\ie}{{\em i.e.}}
\newcommand{\eg}{{\em e.g.}}

\newcommand{\PT}{$\mathcal{PT}$}

\title{Influence of non-Hermitian mode topology on refractive index sensing with plasmonic waveguides}

\author[1,2]{Alessandro Tuniz}
\author[3,4]{Markus A. Schmidt}
\author[1]{Boris T. Kuhlmey}

\affil[1]{Institute of Photonics and Optical Science (IPOS), School of Physics, The University of Sydney, NSW 2006, Australia}
\affil[2]{The University of Sydney Nano Institute (Sydney Nano), The University of Sydney, NSW 2006, Australia}
\affil[3]{Leibniz Institute of Photonic Technology (IPHT Jena), Albert-Einstein-Str. 9, 07745 Jena, Germany}
\affil[4]{Abbe Center of Photonics and Faculty of Physics, Friedrich-Schiller-University Jena, 07743 Jena, Germany}

\affil[*]{Corresponding author: alessandro.tuniz@sydney.edu.au}

\doi{\url{http://dx.doi.org/10.1364/XX.XX.XXXXXX}}

\begin{abstract}
We evaluate the sensing properties of plasmonic waveguide sensors by calculating their resonant transmission spectra in different regions of the non-Hermitian eigenmode space. We elucidate the pitfalls of using modal dispersion calculations in isolation to predict plasmonic sensor performance, which we address by using a simple model accounting for eigenmode excitation and propagation. Our transmission calculations show that resonant wavelength and spectral width crucially depend on the length of the sensing region, so that no single criterion obtained from modal dispersion calculations alone can be used as a proxy for sensitivity. Furthermore, we find that the optimal detection limits occur where directional coupling is supported, where the narrowest spectra occur. Such narrow spectral features can only be measured by filtering out all higher-order modes at the output, e.g., via a single-mode waveguide. Our calculations also confirm a characteristic square root dependence of the eigenmode splitting with respect to the permittivity perturbation at the exceptional point, which we show can be identified through the sensor beat length at resonance. This work provides a convenient framework for designing and characterizing plasmonic waveguide sensors when comparing with experimental measurements.
\end{abstract}

\setboolean{displaycopyright}{true}

\begin{document}

\maketitle

\section{Introduction}
Surface plasmon polariton resonant sensors~\cite{guo2012surface} have found wide-ranging applications, particularly for nanoscale bio-sensing~\cite{chung2011plasmonic}, where they allow label-free optical detection of binding events between molecules (e.g., antibodies and antigens~\cite{coskun2014lensfree}), protein interactions~\cite{beuwer2015stochastic} or exosomes~\cite{im2014label}. The original prism-based geometries such as the Kretschmann~\cite{kretschmann1968radiative} and Otto~\cite{otto1968excitation} configurations are versatile and precise, but rely on free space optics and are thus relatively bulky. Surface plasmon lend themselves to extreme confinement of light, which could be implemented in very small footprint devices, motivating extensive research investigating strategies for  integrating surface plasmon sensors with chip-based~\cite{chamanzar2013hybrid, peyskens2016surface} and fiber-based~\cite{vaiano2016lab, tuniz2018interfacing} circuitry.

\begin{figure*}[t!]
\centering
\includegraphics[width=0.65\textwidth]{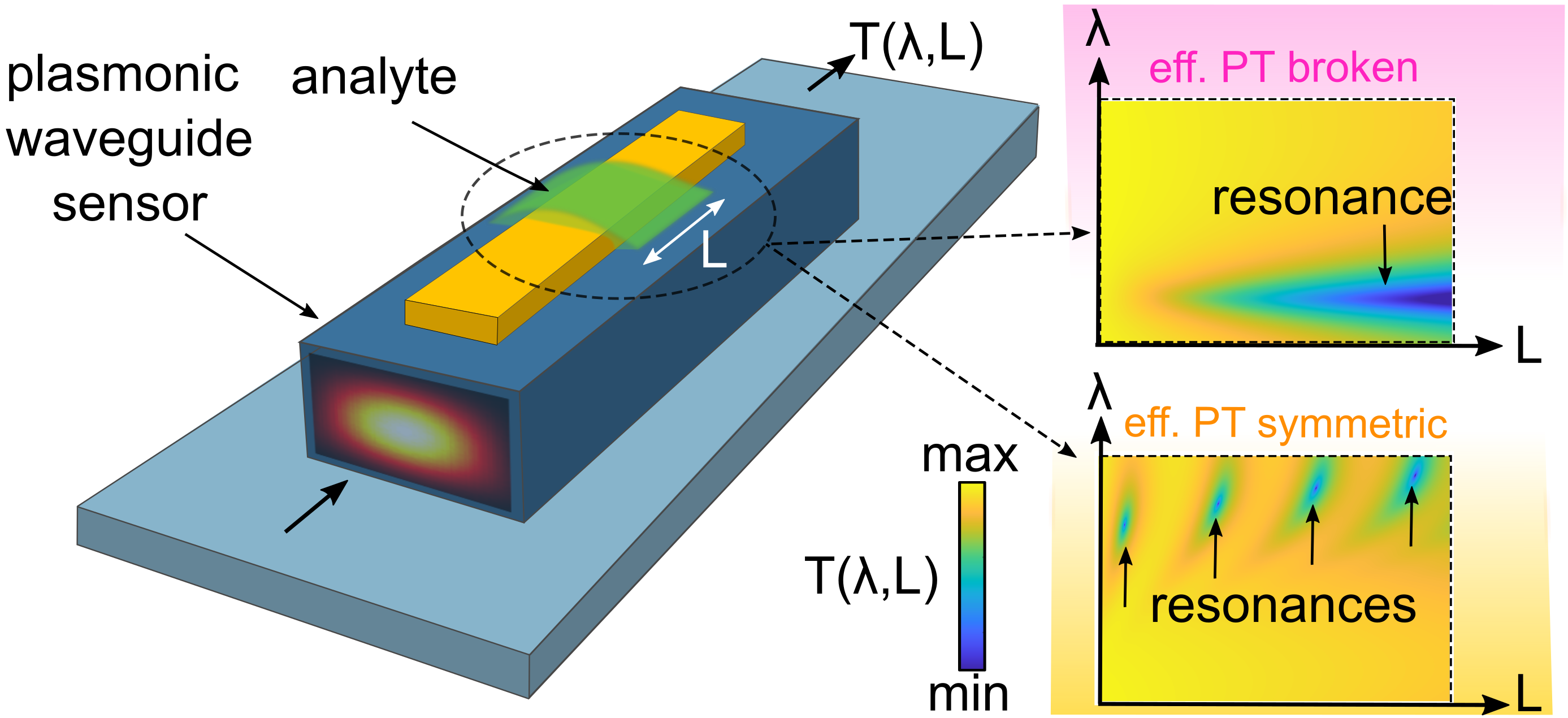}
\caption{Concept schematic overview of the present study. A plasmonic waveguide sensor can be used to identify change in the refractive index of an analyte (blue: dielectric; yellow: metal; green: analyte; interaction length: $L$) by coupling to a mode at the input and measuring the transmission spectrum $T$ at the output. Note that the resonant transmission is a function of both wavelength $\lambda$ and length $L$, can can present remarkably different characteristics depending on whether the system is in the effective parity-time ($\mathcal{PT}$) symmetric, or effective $\mathcal{PT}$ broken regime~\cite{tuniz2019tuning}, each unlocked by changing the refractive index of the analyte.}
\label{fig:fig_concept}
\end{figure*}

Experimental and theoretical work on waveguide-based plasmonic sensors goes back several decades~\cite{harris1995waveguide, homola1999surface}: we refer the reader to Refs.~\cite{caucheteur2015review,klantsataya2017plasmonic, xu2019optical} for a selection of recent reviews. Such devices are commonly composed of rectangular-~\cite{dostalek2001surface, peyskens2016surface}, cylindrical-~\cite{piliarik2003surface, wieduwilt2013optical, wieduwilt2015ultrathin, tuniz2019tuning}, or microstructured-~\cite{wang2009side, rifat2018highly} dielectric waveguides (e.g., composed of silica~\cite{wieduwilt2015ultrathin}, PMMA~\cite{wang2009side}, or silicon nitride~\cite{peyskens2016surface}), adjacent to one- or several- metallic nanostructures (e.g., nanofilms~\cite{wieduwilt2015ultrathin, rifat2018highly}, nanowires~\cite{gu2013metal}, and nanoantennas~\cite{peyskens2016surface}), which are in contact with a region to be sensed (e.g., a liquid~\cite{wieduwilt2015ultrathin}, or a gas~\cite{nau2010hydrogen, caucheteur2016ultrasensitive}). Because dielectric waveguides and plasmonic films typically differ by orders of magnitude in lateral dimensions, it is generally challenging to couple light efficiently between them. One approach is to tailor the geometry of each waveguide such that individual uncoupled propagation constants are equal at a particular wavelength~\cite{degiron2009directional, tuniz2016broadband}, i.e., they are phase matched, analogously to what occurs in sensors that rely on dielectrics alone~\cite{wu2009ultrasensitive, lee2011optofluidic, wu2013performance,lee2014refractive}.  Figure~\ref{fig:fig_concept} (left) shows a concept schametic of an example plasmonic waveguide sensor. After coupling light into the dielectric core, a wavelength-dependent excitation/propagation of the modes in the sensing region occurs, leading to characteristic transmission spectra (Fig.~\ref{fig:fig_concept}, right). In typical sensing schemes, the phase matching wavelength is associated with a local transmission dip due to directional coupling; because this condition is sensitive to the refractive index, shifts in the transmitted spectrum thus contain information on changes in the analyte. Compared to all-dielectric sensors, plasmonic sensors can exploit plasmonic modes with extremely short evanescent tails, which allow for exquisite sensitivity  to small refractive index changes within a few tens of nm~\cite{chung2011plasmonic}. However, because plasmonic systems are inherently lossy (i.e., non-Hermitian~\cite{alaeian2014non, feng2017non}), they feature subtle and counter-intuitive eigenmode topologies~\cite{miri2019exceptional}, whose excitation and propagation is far from trivial~\cite{zhong2018winding}, and thus demand careful consideration.  Figure~\ref{fig:fig_concept} (right) shows different achievable transmission spectra which can result by varying wavelength and interaction length. Depending on the analyte refractive index, such sensors can behave as ``effective'' parity-time ($\mathcal{PT}$) symmetric (EPTS), or effective $\mathcal{PT}$ broken (EPTB) systems~\cite{ozdemir2019parity,tuniz2019tuning}, whose resonant behaviour is markedly different when the analyte length $L$ is changed~\cite{tuniz2019tuning}.  Furthermore, non-Hermitian systems can support exceptional points~\cite{miri2019exceptional} (EPs), which in some experiments can be used for enhanced sensing~\cite{park2020symmetry}; what role EPs play in the specific context of plasmonic waveguide sensors has, to the best of our knowledge, yet to be discussed in detail. Most commonly, experimental reports of plasmonic waveguide sensors are accompanied by  mode simulations to explain the overall measured features, judiciously selecting the mode that dominates the loss spectrum~\cite{wieduwilt2013optical, rifat2018highly}; conversely, several numerical reports are not supported by detailed comparisons with experiments~\cite{liu2017mid, islam2018dual} (often because the designed devices, although realistic, are challenging to fabricate). In such cases sensing performance -- \eg, how much a plasmonic resonance shifts as the analyte index changes -- is deduced from the properties of individual 2D mode calculations, rather than considering the 3D excitation, propagation, and interference of all participating modes. Simulations of plasmonic sensor implementations~\cite{dyshlyuk2018waveguide} using commercially available solvers (FEM, FDTD) often requires fine meshing and large devices, making them computationally demanding and time consuming. While full device transmission spectra have been discussed for selected configurations~\cite{fan2012refractive,dyshlyuk2018waveguide,tuniz2019tuning}, a comprehensive study of how the inferred sensing properties fare against experimentally measurable quantities as a function of the key parameters is still missing. 

\begin{figure*}[t!]
\centering
\includegraphics[width=0.9\textwidth]{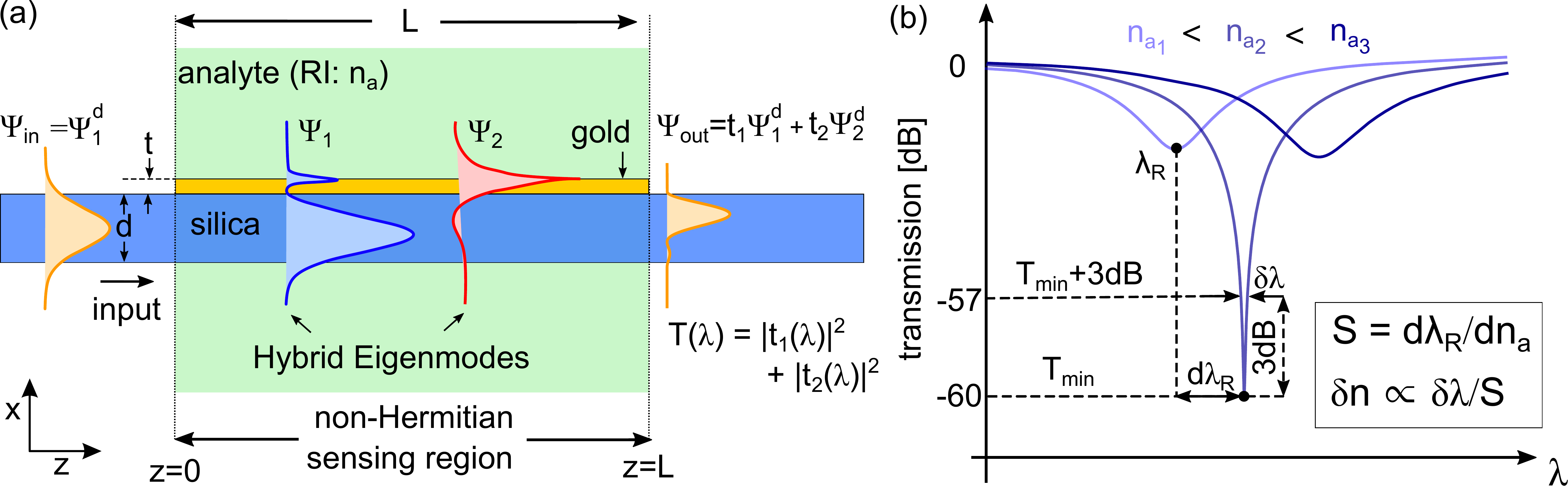}
\caption{Concept schematic of the plasmonic waveguide sensor considered. The fundamental mode input of a dielectric silica waveguide (width: $d$; field: $\psi_{in} = \psi_1^d$) couples to the hybrid eigenmodes (HEMs; blue and red curves: $\psi_{1,2}^{HEM}$)  of a gold-coated region (thickness: $t$), surrounded by a liquid of refractive index $n_a$. The eigenmode excitation and interference over a length $L$ results in a wavelength-dependent transmitted power $T$, which can contain information on changes in $n_a$. The output field is a superposition of the dielectric waveguide eigenmodes, $\psi_{out} = t_1\psi_1^d + t_2\psi_2^d + \ldots$ The sensing region is lossy and thus non-Hermitian. (b) Example $T(\lambda)$ spectra for increasing $n_a$. The shift in resonant wavelength $\lambda_R$ determines the sensitivity $S = d\lambda_R/dn_a$. Each resonance possesses a characteristic 3dB-width $\delta\lambda$ which depends on eigenmode excitation and interference upon propagation. Small changes in $n_a$ can be resolved for small $\delta\lambda$ and large $S$, i.e., the detection limit is $\delta n \propto\delta\lambda/S$~\cite{wu2013performance}.}
\label{fig:fig1}
\end{figure*}

Here we show, with 1D modes undergoing 2D propagation, that full transmission characteristics of non-Hermitian plasmonic waveguide sensors can be reproduced with a straightforward eigenmode model~\cite{tuniz2016broadband} that relies on modal calculations. This model crucially requires takes into account both coupling in and out of the device, as well as propagation over a finite length. Our results are  validated by full vector finite element method calculations (COMSOL). These fast computations enable us to calculate the full transmission characteristics, such as the extinction ratio, resonance width, and sensitivity, as a function of all key parameters, including wavelength, interaction length, and analyte index. This approach allows a direct comparison of full device performance (\ie, which considers mode excitation and propagation) with commonly used approaches that use modal calculations alone (\ie, which don't consider excitation and propagation). We show that many proxies extracted from mode calculations or coupled more theory alone, such as phase matching wavelength, are not representative of a device's performance, and can provide misleading sensitivity and detection limit values. Our study reveals a number of additional key features:  (i) the detection limit is a property of a specific device, and crucially depends on the physical sensor length; (ii) the lowest detection limits occur in regions where directional coupling occurs, as a result of beating between hybrid modes which produce narrow spectral widths; (iii) single-mode filtering at the multi-mode sensor output is fundamental for achieving such a narrow spectral feature; (iv) the high sensitivity at the exceptional point can be seen in the modal beating of the hybrid eigenmodes in the plasmonic region.  These results  can be immediately adapted to more realistic systems formed by 2D modes undergoing 3D propagation~\cite{degiron2009directional}. 

\section{THEORY}

\subsection*{Principle of Operation}

We first consider the plasmonic waveguide sensor shown in Fig.~\ref{fig:fig1}(a), where modes propagate in $z$, and different materials are distributed in $x$. Here we will limit ourselves to the analysis of a specific pure 2D geometry, which was designed to exemplify the different topological situations of eigenmodes for analytes with refractive index of aqueous solutions, as well as the various sensor performance and design considerations one can encounter. All waveguides are assumed to be infinite in $y$, providing physical insight into coupling mechanisms over a large parameter space and relatively fast calculation times. 
The system is here formed by a 1D slab of SiO$_2$ (waveguide width: $d = 2\,{\mu {\rm m}}$), coated with a finite length gold nano-film on one side (thickness t= 30\,nm), and surrounded by an analyte (RI: $n_a=1.32-1.42$) elsewhere.  This dielectric waveguide is multimode, supporting 3--6 modes between 400--800\,nm. While  we will only excite the fundamental mode at the input, having multiple modes at the output is of importance for our later analysis. The relevant plasmonic mode here is the long-range surface plasmon~\cite{berini2009long}, whose real part can be designed to match with the fundamental mode of the dielectric waveguide. We choose this geometry as a starting point of the discussion for two reasons; firstly, because it most closely resembles commonly-used fiber-based structures~\cite{klantsataya2017plasmonic}; secondly, because long-range plasmons have a cutoff wavelength~\cite{burke1986surface} that is highly sensitive to the environment, and which can also be harnessed for enhancing sensors. For this gold film thickness, the (cutoff-free) short-range surface plasmon's effective index and loss are too high to couple to the dielectric mode, and are considered separately for a suitably modified geometry in Section~\ref{sec:SRSPP}.
The interaction length $L$ corresponds to the length where the gold film and the analyte overlap. 
The modes in the sensing section are excited by the fundamental mode of the silica slab at $z=0$. At $x=L$, the superposition of modes of the sensor in turn excites modes of the silica slab. Experimentally, measurements consider the resulting transmitted wavelength-dependent intensity, shown schematically in Fig.~\ref{fig:fig1}(b). These measurements are characterized by a transmission minimum $T_{min}$, a resonant wavelength $\lambda_R$, and a spectral width $\delta\lambda$, all of which depend on $n_a$.

There are two dominant interpretation of  the cause of the resonant dip in waveguide  plasmonic sensors. The first is that the dominant mode inside the sensor is lossy, with  a loss peak that depends on the analyte's index. In that interpretation, the wavelength-dependent loss of this dominant mode is measured. The second is that {\em two} modes are excited in the sensing section, and the output results from their interference - in short, that the plasmonic sensor acts as directional coupler, with the phase matching wavelength being dependent on the analyte index. In this paper we will show in which circumstances each of these interpretations is correct, their limitations, and how to use their understanding to optimize sensing performance.

\subsection*{Sensitivity and Detection Limit}
Figure~\ref{fig:fig1}(b) shows the main features of typical transmission spectra as $n_a$ increases. The position of the resonance wavelength $\lambda_R$ is a function of $n_a$, and the sensor's sensitivity $S$ is defined by the shift of resonant
wavelength per change in analyte RI,

\begin{equation}
S = \frac{d\lambda_R}{dn_a}.
\label{eq:S}
\end{equation}

A refractive index sensor's overall performance is best characterized by its detection limit (DL), which is the smallest \emph{detectable} change in RI $\delta n$, which generally depends on a specific user's experimental configuration. Wu {\em et al.}~\cite{wu2013performance} extended White and Fan's~\cite{white2008performance} heuristic formula for the detection limit in the context of measurements which rely on resonant transmission spectra, most relevant for the present case,

\begin{equation}
\delta n \approx \frac{1}{S}\frac{\delta\lambda}{1.5({\rm SNR})^{0.25}},
\label{eq:dn}
\end{equation}
where SNR is the signal-to-noise ratio in linear units near the transmission minimum $T_{min}$, and $\delta\lambda$ is the spectral width at twice the minimum transmission, i.e., the $T_{min} +3\,{\rm dB}$ limit as shown in Fig.~\ref{fig:fig1}(b). The smallest detectable $\delta n$ would stem from a combination of narrow resonance width, high sensitivity, and low instrument noise. Since the SNR is dictated by the instrument, for the remainder of this work we shall consider only $S$ and $\delta n$.  In general, a high $S$ and small $\delta \lambda$ do not occur in the same region of parameter space. While $S$ could be inferred from modal calculations (in lossless dielectric directional couplers for example, it can be extracted from the shifts in the phase matching wavelength~\cite{wu2009ultrasensitive,lee2011optofluidic, wu2013performance, pumpe2017monolithic}), $\delta \lambda$ also depends on the interaction length $L$, and thus requires that the modes' excitation and propagation through the sensor be considered in detail. In lossy systems, even calculating $S$ can be difficult, as losses and interference compete, so that the the minimum transmission does not necessarily occur at the phase matching wavelength or at the wavelength of maximum loss for any one mode.

\begin{figure*}[t!]
\centering
\includegraphics[width=1\textwidth]{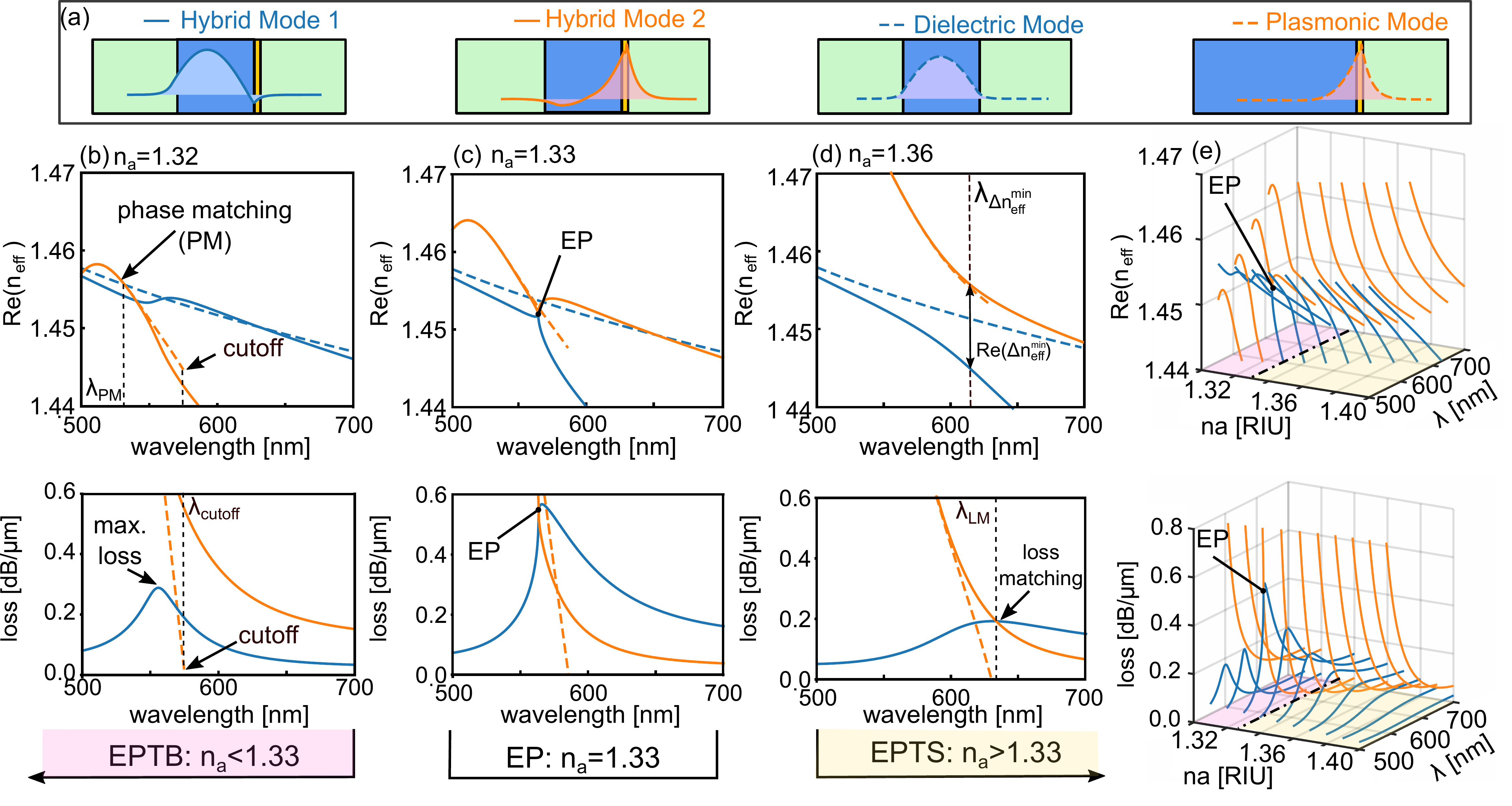}
\caption{(a) Summary schematic of relevant modes. Solid blue/orange curves: hybrid modes of a silica waveguide of finite width in contact with a thin gold film. The blue dashed curve corresponds to an equivalent system without gold film (dielectric mode), and orange dashed curve corresponds to an equivalent system with infinite silica width (plasmonic mode). The associated $\Re e(n_{\rm eff}$) (top row) and loss (in  dB/{$\mu$\rm{m}}, bottom row), as  a function  of wavelength, are shown for (b) $n_a = 1.32$, (c) $n_a = 1.33$, and (d) $n_a = 1.36$. Note the transition from the crossing- to anti-crossing- of $\Re e(n_{\rm eff})$ (and vice versa for the loss via $\Im m(n_{\rm eff})$) via the exceptional point (EP). Also shown are different criteria used in the literature for inferring where plasmonic resonances occur: ``phase matching'' (where the real parts of the dielectric- and plasmonic- eigenmodes cross~\cite{wu2009ultrasensitive}), a maximum loss region (where the  loss of the dielectric-like hybrid mode is maximum~\cite{wieduwilt2013optical}), and a ``loss matching'' region~\cite{zhang2008dependence} (where the imaginary parts of the hybrid eigenmodes cross). The wavelength $\lambda_{\Delta n_{\rm eff}^{\rm min}}$ of minimum effective index difference is also indicative of strong coupling~\cite{taras2021shortcuts}. In the present configuration and near resonance, eigenvalues coalesce at the EP when $n_a=1.33$, $n_a<1.33$ supports \emph{effective $\mathcal{PT}$-broken} (EPTB) modes,  $n_a>1.33$ supports \emph{effective $\mathcal{PT}$-symmetric} (EPTS) modes. (e) Detailed 3D plot of $\Re e(n_{\rm eff}$) (top) and loss (bottom row), as  a function  of wavelength and $n_a$.
}
\label{fig:fig2}
\end{figure*}

\subsection*{Modes}

We begin by considering all relevant bounded eigenmodes shown in the Fig.~\ref{fig:fig1}(a) schematic.  The propagation constants, as well as the electric- and magnetic- fields of each mode are obtained by numerically solving a complex transcendental dispersion equation resulting from enforcing boundary conditions between the layers~\cite{burke1986surface}. The material dispersion for silica~\cite{malitson1965interspecimen} and gold (Drude model in Ref.~\cite{rakic1998optical}) are taken into account, and the analyte index is taken as a wavelength-independent constant as labelled. Figure~\ref{fig:fig2}(a) shows a schematic of the modes considered. 
The two hybrid modes in the sensing region of Fig.~\ref{fig:fig1}(a) (propagation constants: $\beta_{i=1,2}$) are shown as solid lines. For comparison, the  isolated (uncoupled) eigenmodes -- supported by an equivalent dielectric waveguide without a gold film (propagation constant: $\beta_d$), or by an equivalent gold nanofilm sandwiched between silica/analyte on each side -- are shown as dashed lines. 

Figure~\ref{fig:fig2}(b),(c),(d) show the associated eigenmodes' dispersion curves ($n_{{\rm eff},i} = \beta_i/k_0$, $k_0 = 2\pi/\lambda$), for three values of  $n_a$. To aid physical intuition, we show each mode's real part $\Re e(n_{{\rm eff},i})$ (top of Fig.~\ref{fig:fig2}(b-d)) and loss $\alpha_i$ (in dB/$\mu {\rm m}$, bottom of Fig.~\ref{fig:fig2}(b-d)), which is related to the imaginary part of $\Im m(n_{\rm eff,i})$ via

\begin{equation}
\alpha_i [{\rm dB}/\mu{\rm m}] = 10 \log _{10}\{\exp[2\Im m (\beta_{i}) \times 1\mu {\rm m}]\}.
\end{equation}
Figure~\ref{fig:fig2}(e) also shows a detailed 3D plot of the dispersion of the hybrid eigenmodes as a function of $n_a$. Note in particular the transition between a regime where $\Re e(n_{\rm eff})$ cross, and $\Im m(n_{\rm eff})$ anti-cross [Fig.~\ref{fig:fig2}(b)], to a regime where $\Re e(n_{\rm eff})$ anti-cross and $\Im m(n_{\rm eff})$ cross [Fig.~\ref{fig:fig2}(d)], separated by an exceptional point (EP) [Fig.~\ref{fig:fig2}(c)] where the propagation constants coalesce, with equal real and imaginary parts for both hybrid modes~\cite{miri2019exceptional}. This kind of transition is characteristic of coupled non-Hermitian systems, and is frequently encountered when designing plasmonic waveguide sensors~\cite{islam2018dual, tuniz2019tuning}. 
The transition through the exceptional point is similar to those in parity-time ($\mathcal{PT}$) symmetric systems~\cite{ozdemir2019parity}, for example in situations including perfectly balanced optical gain and losses~\cite{ozdemir2019parity}. In such $\mathcal{PT}$ symmetric situations, the eigenvalues can either be real and follow parity-time symmetry, or form complex conjugate pairs and thus break $\mathcal{PT}$ symmetry. These two regimes are separated by the EP, where the eigenvalues coalesce. Structures with a global net loss, e.g., plasmonic~\cite{tuniz2019tuning, park2020symmetry} and leaky~\cite{khurgin2021emulating} waveguides, share many of the same features as non-Hermitian systems with no net loss: the main difference is that eigenvalues are shifted along the positive imaginary axis with respect to the perfectly loss-balanced case. Coupled lossy systems can thus be more rigorously classified as having eigenvalues that are either \emph{effective $\mathcal{PT}$-symmetric} (EPTS) or \emph{effective $\mathcal{PT}$-broken} (EPTB), as labelled in Fig.\ref{fig:fig2}(a). To simplify the discussion and analogously to an earlier report~\cite{tuniz2019tuning}, we point out that in the present configuration and near regions where plasmonic resonances are measured, eigenvalues coalesce at the EP when $n_a=n_{EP}=1.33$, $n_a<1.33$ supports EPTB modes,  $n_a>1.33$ supports EPTS modes, as highlighted by Fig.~\ref{fig:fig2}(b)--(e). In the EPTB regime near the resonance, both hybrid eigenmodes have field distributions with sizeable overlap with the dielectric core region,  and have comparable loss: they are both excited by the incoming dielectric mode, and their interference leads to directional coupling and energy exchange between the dielectric waveguide and metal film~\cite{degiron2009directional,tuniz2016broadband}. In the EPTS regime, the incoming dielectric mode predominantly excites the low-loss hybrid mode, effectively leading to monotonic exponential decay resulting from the transmission of the excited lossy mode~\cite{tuniz2019tuning}.

\begin{figure*}[t!]
\centering
\includegraphics[width=\textwidth]{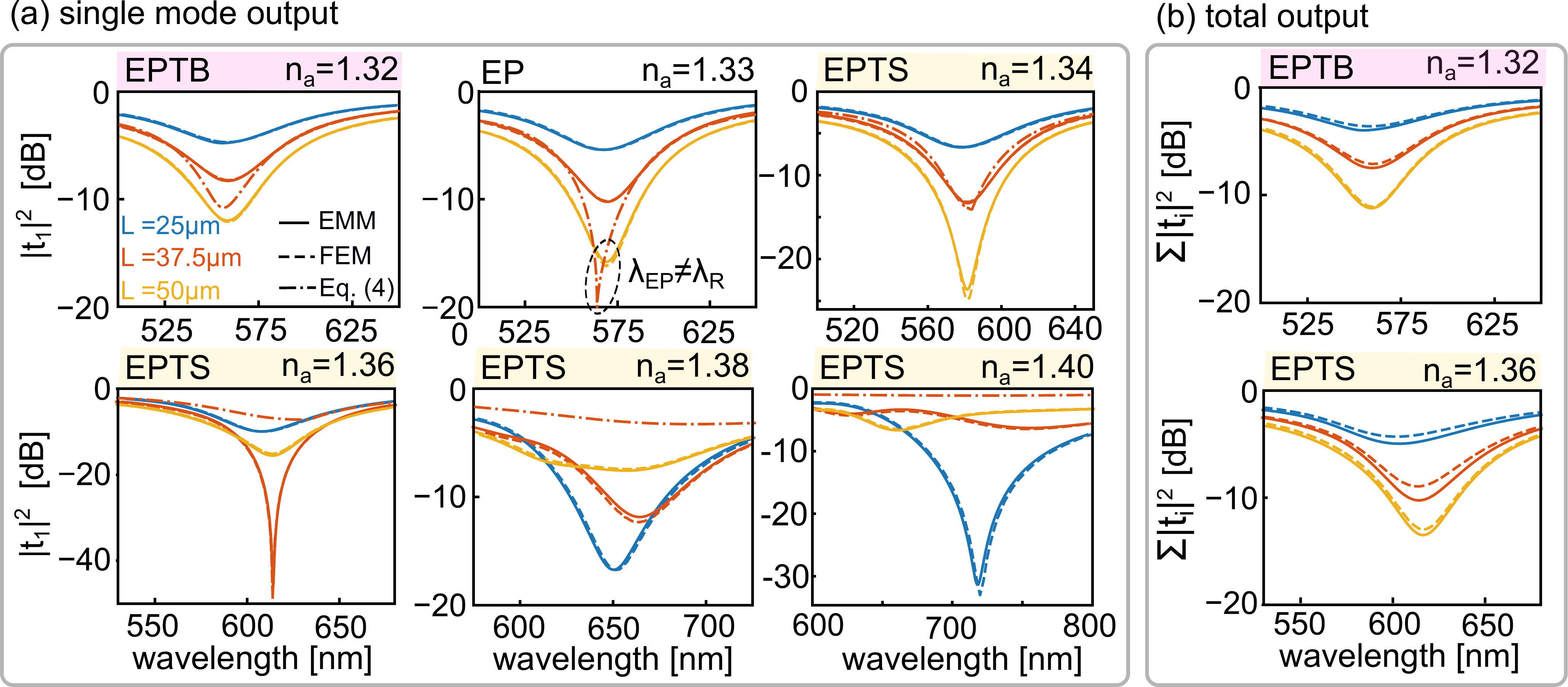}
\caption{(a) Spectral distribution of power in the fundamental dielectric waveguide mode at output as a function of wavelength $T(\lambda) = |t_1(\lambda)|^2$, for the four analyte indices of Fig.~\ref{fig:fig2} as labelled, for $L=25\,\mu{\rm m}$, $L=37.5\,\mu{\rm m}$, and $L=50\,\mu{\rm m}$ (blue, orange and yellow respectively). Dashed-dotted lines compute Eq.~\ref{eq:T} for  one example length ($L=37.5\,\mu{\rm m}$).
(b) Total power in the waveguide at output as a function of wavelength, i.e., $T(\lambda) = \Sigma_i |t_i(\lambda)|^2$. Solid lines correspond to calculations performed via the eigenmode method (EMM) including mode excitation and propagation, and dashed lines correspond to full vector finite element method (FEM) calculations (COMSOL).}
\label{fig:fig3}
\end{figure*}

One important question that arises when designing plasmonic refractive index sensors is the following: How can the location and shift of the resonant wavelength $\lambda_R$ be inferred from the modal calculations of Fig.~\ref{fig:fig2}? Various approaches may be found in the literature, as a result of the many possible choices available, which result from the transition between EPTS and EPTB regions near resonance. Inspecting the loss of hybrid mode 1 in the EPTB region (e.g., Fig.~\ref{fig:fig2}(b), $n_a=1.32$) might suggest a simple interpretation: the fundamental mode of the dielectric waveguide couples only to the lowest loss mode, the resonant transmission $T(\lambda)$ (in dB) can be computed as

\begin{equation}
T(\lambda) = {\rm min} [\alpha_1, \alpha_2] \times  L,
\label{eq:T}
\end{equation}
where $\alpha_i$ is the loss of hybrid Mode $i$ (in ${\rm dB}/\mu{\rm m}$) and $L$ is the interaction length (in $\mu{\rm m}$). In other words, the loss is computed from the minimum of the two hybrid mode loss curves of Fig.~\ref{fig:fig2}(c), and both $\lambda_R$ and $\delta \lambda$ follow immediately from $\alpha_1$, and are independent of $L$. $T_{min}$ thus occurs where the loss of this curve is maximum, as highlighted by the label ``max. loss'' in Fig.~\ref{fig:fig2}(c). We will show that this approach is adequate {\em in the EPTB} region, but many reports extend this reasoning to the EP and EPTS regions -- {\em i.e.}, by computing the loss from the minimum of the two hybrid mode loss curves in Figs.~\ref{fig:fig2}(c) at each $n_a$ -- without accounting for mode excitation, propagation, and interference. This reasoning would imply that the resonance minimum $T_{min}$ occurs at the wavelength where the two loss curves of the hybrid eigenmodes in Figs.~\ref{fig:fig2}(c) intersect, {\em i.e.}, the ``loss matching'' wavelength $\lambda_{LM}$~\cite{zhang2008dependence}, indicated by ``LM'' in Fig.~\ref{fig:fig2}(c).
One other choice that can be borrowed from the dielectric waveguide literature to predict $\lambda_R$ is the ``phase-matching'' (PM) wavelength $\lambda_{PM}$, {\em i.e.}, where the real parts of the uncoupled effective indices cross, as indicated by ``PM'' in Fig.~\ref{fig:fig2}(b). The PM wavelength would be a reasonable estimate of the resonant wavelength of the directional coupler consisting of the plasmonic and dielectric waveguides, if these two waveguides were {\em weakly} coupled. However, plasmonic refractive index sensors use waveguides that are often in close proximity, invalidating this approximation. Furthermore, because the plasmonic guide has a cutoff, the phase matching point ceases to exist at higher analyte indices. In the EPTS region, where directional coupling occurs, one alternative choice is the wavelength $\lambda_{\Delta n_{\rm eff}^{\rm min}}$ at the minimum effective index difference, \ie, where $\Re e(\Delta n_{\rm eff}^{\rm min})  = min[\Re e( n_{\rm eff,2} - n_{\rm eff,1})]$, as highlighted in Fig.~\ref{fig:fig2}(d). This wavelength informs where coupling between adjacent waveguides is strongest~\cite{taras2021shortcuts}, and resonant energy transfer is most efficient. 
Finally, another proxy for the resonant wavelength one can  use is the cutoff wavelength of the plasmonic mode  $\lambda_{\rm cutoff}$ as indicated by ``cutoff'' in Fig.~\ref{fig:fig2}(c). This is perhaps less used in the plasmonic literature, but has been used to successfully analyse the high sensitivity of all-dielectric photonic crystal fiber sensors with liquid analyte satellites~\cite{wu2009ultrasensitive}. Note that the dependence of these parameters on $n_a$, should it correlate with $\lambda_R$, would only provide information on $S$ via \eqref{eq:S}, but that $\delta\lambda$ and $\delta n$ require knowledge of the full transmission spectrum - which generally cannot be obtained simply from looking at modal dispersion curves alone. 

\subsection*{Propagation}

We now evaluate the transmission spectra associated with the waveguide configuration shown in Fig.~\ref{fig:fig1}, accounting for the modes' excitation and propagation along the device length $L$ as $n_a$ is varied. Here, we always consider the case where the input mode is the fundamental mode of the isolated dielectric waveguide (\ie, a silica core and an analyte cladding), corresponding to typical experimental conditions~\cite{klantsataya2017plasmonic}. The fundamental mode couples to the two hybrid eigenmodes (EM) of the plasmonic sensor, which then propagate through the sensing region before exciting the modes of the output dielectric waveguide. Although this simple model does not account for reflections and scattering at the boundary between the dielectric- and hybrid- waveguide, these have a negligible impact on the overall transmission spectra at the configurations considered, as we discuss below. For 1D modes propagating along $z$, the total electric and magnetic field in the sensing region can be written as a superposition of its supported eigenmodes,

\begin{equation}
\mathbf{E}(x,z) =  a_1  \mathbf{E}_1(x) \exp(i\beta_1 z) 
+ a_2 \mathbf{E}_2(x) \exp(i\beta_2 z),
\label{eq:EM1}
\end{equation}
\begin{equation}
\mathbf{H}(x,z) =  a_1  \mathbf{H}_1(x) \exp(i\beta_1 z) 
+ a_2 \mathbf{H}_2(x) \exp(i\beta_2 z), 
\label{eq:EM2}
\end{equation}
where the subscript $i = 1,2$ labels the two hybrid EMs in the sensing region. Here we use an implicit $\exp(-i\omega t)$ time dependence, and the convention that actual fields are the real parts of the complex fields. $\mathbf{H}$ and $\mathbf{E}$ are the transverse total magnetic and electric field vectors at any point in the plasmonic waveguide, respectively, $\mathbf{H}_i$ and $\mathbf{E}_i$ are the magnetic and electric field distributions of the hybrid EMs, $\beta_i = \beta_i^R + i\beta_i^I$ are their propagation constants, and $a_i$ are the modal amplitudes which determine the contribution of the respective EMs to the total field. Strictly speaking, the above expansions are approximate, since contributions from higher order modes and radiation modes have been neglected, but we shall be verifying their validity through comparisons with full field finite element calculations.  Following the convention used in~\cite{snyder2012optical}, modal fields are normalized so that

\begin{equation}
\frac{1}{2} \int \left[\mathbf{E}_i(x) \times \mathbf{H}_j(x) \right] \,dx = \delta_{i,j}.
\label{eq:norm}
\end{equation}
This normalization does not use the complex conjugate of $\mathbf H$, since the latter is problematic in waveguides including material losses~\cite{snyder2012optical}. Note, however, that this normalization is equivalent to the complex conjugated version for purely lossless waveguides.

Modes can always be normalized to satisfy \eqref{eq:norm}, which can then be used to calculate the complex modal amplitudes at input, leading to
\begin{equation}
a_i = \frac{1}{2} \int \left[\mathbf{E}_i(x) \times \mathbf{H}_{in}(x) \right] \,dx,
\end{equation}
where in this case $\mathbf{H}_{in}(x) = \mathbf{H}_1^d(x)$ is the magnetic field of the fundamental mode of the dielectric waveguide, corresponding to the input ($z=0$), as shown in Fig.~\ref{fig:fig1}(a). 
Note that because~\eqref{eq:norm} is unconjugated and the waveguide includes losses, the sum of $|a_i|^2$ is {\em not} equal to the total power carried in the waveguide.
At the output boundary between the plasmonic sensor and the dielectric waveguide ($z=L$), the field is a superposition of the eigenmodes of the plasmonic waveguide, \ie, $\mathbf{H}_{out}(x) = \mathbf{H}(x,L)$ via \eqref{eq:EM2}. The transmitted amplitude is then obtained by projecting this field onto the modes of the silica waveguide via

\begin{equation}
t_i = \frac{1}{2} \int\left[\mathbf{E}_i^d(x) \times \mathbf{H}_{out}(x) \right]  \,dx,
\end{equation}
where $\mathbf{E}_i^d$ is the electric field distributions of each dielectric waveguide eigenmode. 
In the purely lossless dielectric waveguide, the normalization~\eqref{eq:norm} is identical to power normalization, and thus the total transmission is given by $T(\lambda) = \sum_i |t_i(\lambda)|^2$; in cases where the output is a single mode waveguide, $T = |t_1|^2$.

\begin{figure*}[t!]
\centering
\includegraphics[width=\textwidth]{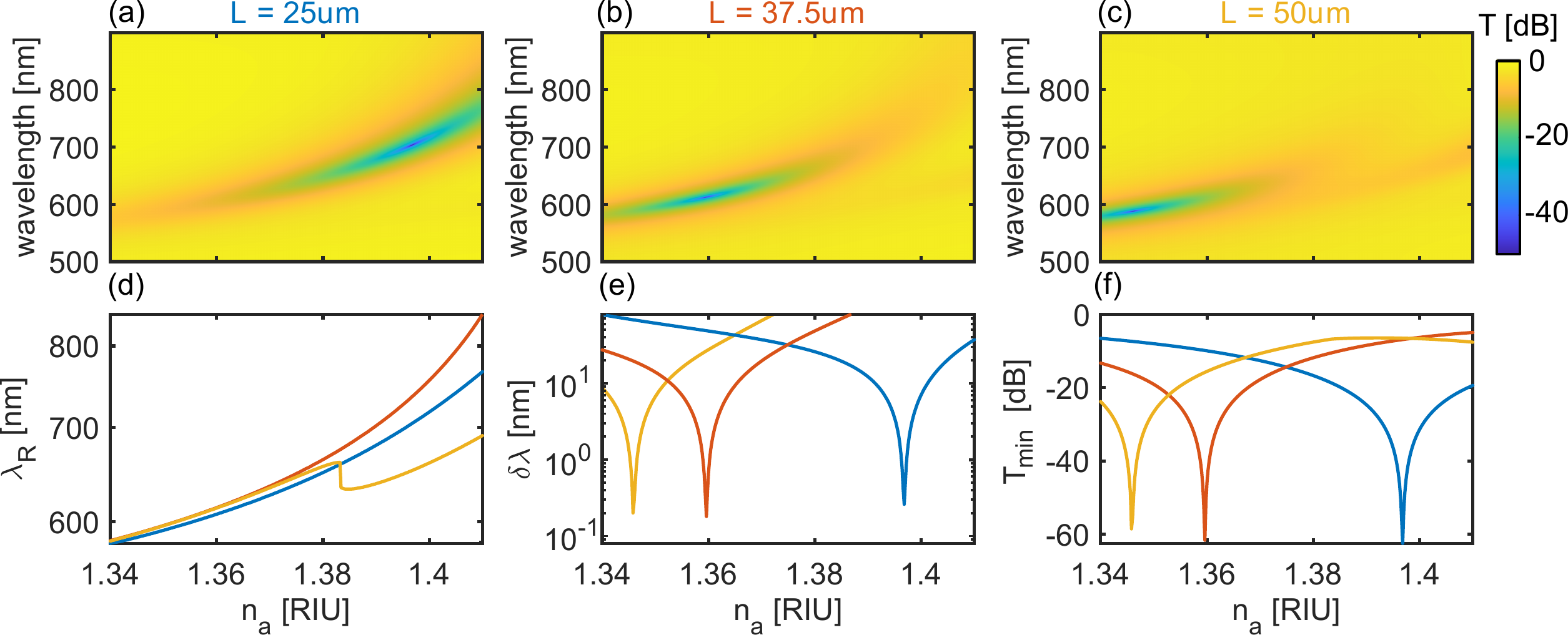}
\caption{Calculated transmission spectra (single mode output) as a function of $n_a$ and $\lambda$ for different values of $L$: (a) $L=25\,\mu{\rm m}$ (blue); (b) $L=37.5\,\mu{\rm m}$ (orange) and (c) $L=50\,\mu{\rm m}$ (yellow). (d) Associated $\lambda_R$, (e) $\delta\lambda$, and (f) $T_{\rm min}$. Line colours in (d)--(f) correspond to that of the $L$ values labelled in (a)--(c). All plots are in the EPTS regime, where the smallest $\delta \lambda$ is achievable due to directional coupling.}
\label{fig:fig4}
\end{figure*}

\begin{figure}[b!]
\centering
\includegraphics[width=0.4\textwidth]{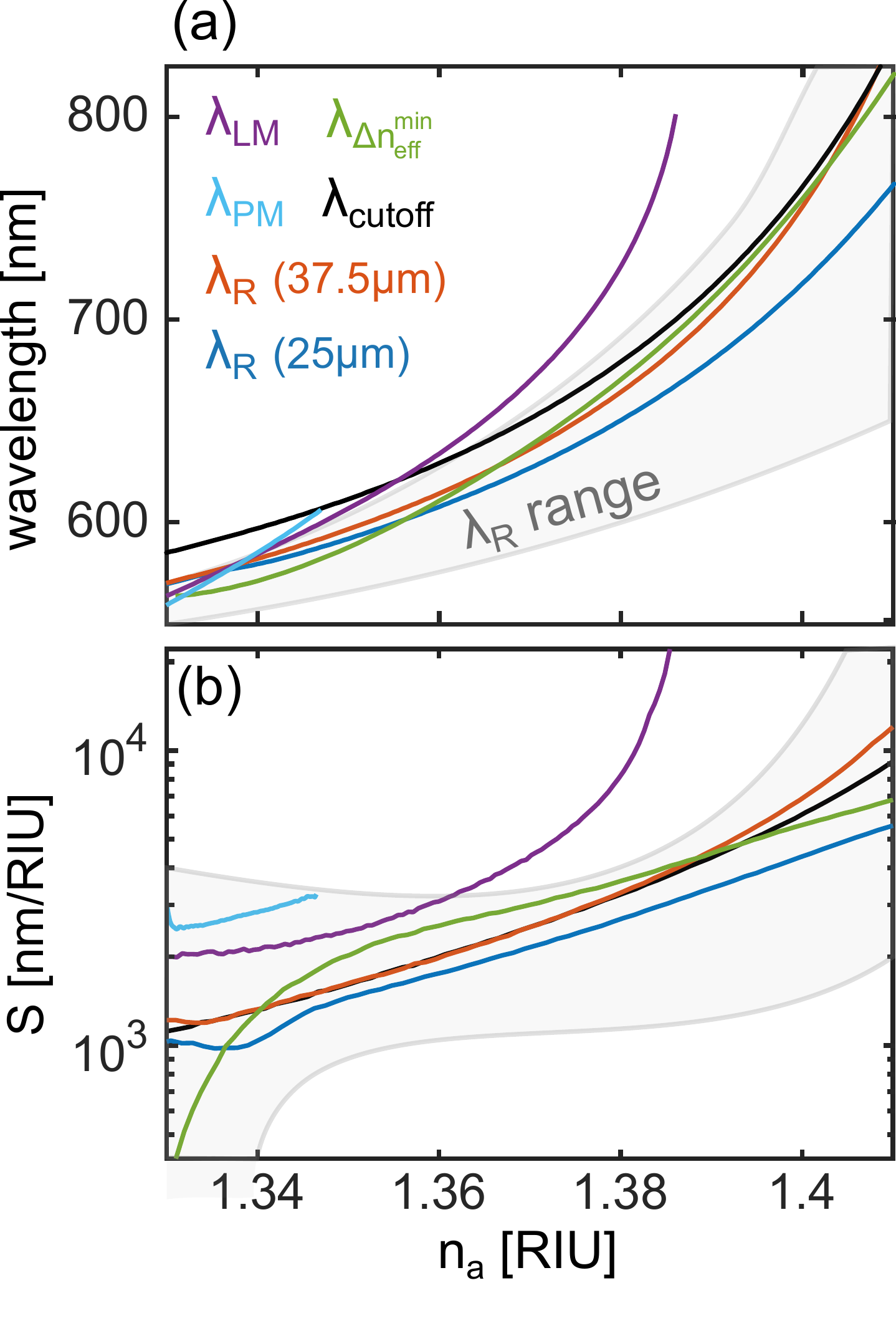}
\caption{(a) Resonant wavelength $\lambda_R$ as a function of analyte index $n_a$, obtained from full transmission calculations, for $L=25\,\mu{\rm m}$ (dark blue) and $L=37\,\mu{\rm m}$ (orange), corresponding to Fig.~\ref{fig:fig4}(d). The grey shaded region shows the entire range of possible $\lambda_R$ for $L=10-200\,\mu{\rm m}$. Cutoff wavelength (black), phase matching wavelength (light blue), loss-matching wavelength (purple), and minimum effective index (green), as a function of analyte index, obtained from mode calculations, and as defined in Fig.~\ref{fig:fig2}. (b) Corresponding sensitivity $S=d\lambda_R/dn_a$ for each curve in (a) (log-scale). The grey shaded region shows the entire range of possible $S$ using $\lambda_R$ for $L=10-2000\,\mu{\rm m}$.}
\label{fig:fig5}
\end{figure}

We calculate the wavelength-dependent modal amplitudes $a_i$, the propagation constants $\beta_i$, and all relevant fields, which in combination are used to calculate the amplitudes $t_i$ at the output interface of the plasmonic sensor region. Figure~\ref{fig:fig3}(a) shows the intensity transmitted by the fundamental mode (\ie, $T(\lambda) = |t_1(\lambda)|^2$) for three examples of interaction lengths $L=25\,\mu{\rm m}$ (solid blue curves), $L=37.5\,\mu{\rm m}$ (solid red curves), and $L=50\,\mu{\rm m}$ (solid orange curves), for increasing analyte index as labelled, associated with the mode calculations of  Fig.~\ref{fig:fig2}.
Before proceeding with a more detailed analysis, we can already note a number of important features. In the EPTB region ($n_a = 1.32$), the transmission spectrum monotonically and exponentially decreases with increasing length, and the resonant wavelength remains nominally unchanged. The transmission spectrum near the EP ($n_a=1.33$) exhibits similar overall features, but with a larger overall loss. Note that the EP wavelength ($\lambda_{\rm EP}=564\,{\rm nm}$) does not correspond to the resonant wavelength at $T_{min}$ ($\lambda_R = 570\,{\rm nm}$), as highlighted by the dashed circle in Fig.~\ref{fig:fig3}(a). Further increasing the analyte index to $n_a=1.36$ leads to a transmission spectrum that oscillates with propagation length due to interference between the excited eigenmodes, noting that here $T_{min}$ and $\delta\lambda$ are smallest at at the intermediate length ($L=37.5\,\mu{\rm m}$). 
One immediate consequence is that the lowest detection limits $\delta n$ as per \eqref{eq:dn} will occur in EPTS regions where directional coupling is supported, and for specific analyte lengths. Finally, at larger analyte indeces  ($n_a=1.38$ and $n_a=1.40$), the resonant transmission spectrum becomes increasingly complicated: $\lambda_R$ and $\delta \lambda$ depend on length, and the eigenmodes' excitation, propagation, and loss contribute to the overall transmission in ways that are challenging to predict by inspecting the modal dispersion alone. This is particularly important from an applications standpoint, since higher analyte indices fundamentally yield larger sensitivities~\cite{wu2009ultrasensitive}. 
Finally, we note that in contrast to the eigenmode theory presented here, a perturbative coupled mode analysis~\cite{chuang1987coupled}, while providing insight into overall non-Hermitian behaviour, cannot rigorously be applied to the present case, because (i) the isolated plasmonic mode cuts off before longer wavelengths regions where hybrid modes are supported, and where the highest sensitivities can be reached and (ii) because the two waveguides are physically connected, and thus no evanescent field is present. Analogously to previous analyses of similar systems, the eigenmode method presented here is the most straightforward method to compute the mode properties and transmitted spectra.

 Figure.~\ref{fig:fig3} also includes a comparison of the  transmission spectra obtained via the eigenmode (EM) method described above with full vector finite element method (FEM) simulations (COMSOL),  shown as dashed lines. A port boundary condition at the input ensures that only the fundamental TM mode of the waveguide is excited. A port at the output provides the option of considering either the amount of power in the fundamental mode (Fig.~\ref{fig:fig3}(a)), or the \emph{total} total power transmitted by the waveguide (Fig.~\ref{fig:fig3}(b)). Perfectly matched layers at all other boundaries suppress any reflections in the simulation volume. We find an excellent agreement between our EM method and the FEM calculations in all cases, provided meshing of the gold region is fine enough $(<10\,{\rm nm})$. The EM method accurately predicts the transmission spectra with computations which are orders of magnitude faster than FEM, making large parameter sweeps more practical. Our FEM calculations indicate that the reflected power back into the input waveguide is $<0.03\%$ for $L>10\,\mu{\rm m}$ for the analyte indices and wavelengths discussed, which \emph{a posteriori} validates our EM model assumptions in the sensor configurations considered.

For completeness, Fig.~\ref{fig:fig3}(a) shows the computed transmission spectrum from \eqref{eq:T}, in the example case of $L=37.5\,\mu{\rm m}$, as a dash-dotted line. While \eqref{eq:T} predicts the transmission in regions where the isolated mode are in the topological vicinity of the coupled modes (i.e., off-resonance), and may be considered adequate in finding an approximate location for $\lambda_R$ for EPTB regions, this model fails to be accurate at resonance for all cases, and fails entirely in the EPTS region. 

Figure~\ref{fig:fig3}(b) shows the same calculations when considering the \emph{total} power transmitted at output, i.e., $T(\lambda) = \Sigma_i |t_i(\lambda)|^2$ for two representative values of $n_a$ in the EPTB and EPTS regions. While this only marginally impacts the EPTB region ($n_a = 1.32$), this is more significant in the EBTS region where directional coupling is supported, because the output power is distributed amongst the available dielectric waveguides modes, increasing $T_{min}$ and decreasing $\delta \lambda$. It immediately follows that, in order to minimize $\delta \lambda$ and thus $\delta n$, it is important to filter out highers order modes at output, for example by splicing the sensor with a single mode fiber~\cite{vaiano2016lab}. Henceforth, we shall therefore  consider such filtering to be implemented and only consider power in the fundamental mode.

\section{DISCUSSION}

The power of the EM model is that it allows us to rapidly obtain resonance wavelength, spectral width, and extinction ratio from rapid calculations of the realistic transmission spectrum $T(\lambda)$ of the full device. Having noted the salient features of plasmonic sensors at different non-Hermitian regimes, and having validated our EM model, we now quantify the key parameters $S$, $\delta\lambda$ and $T_{min}$ in detail, particularly with regards to their dependence on $L$ and $n_a$.

\begin{figure*}[t!]
\centering
\includegraphics[width=1\textwidth]{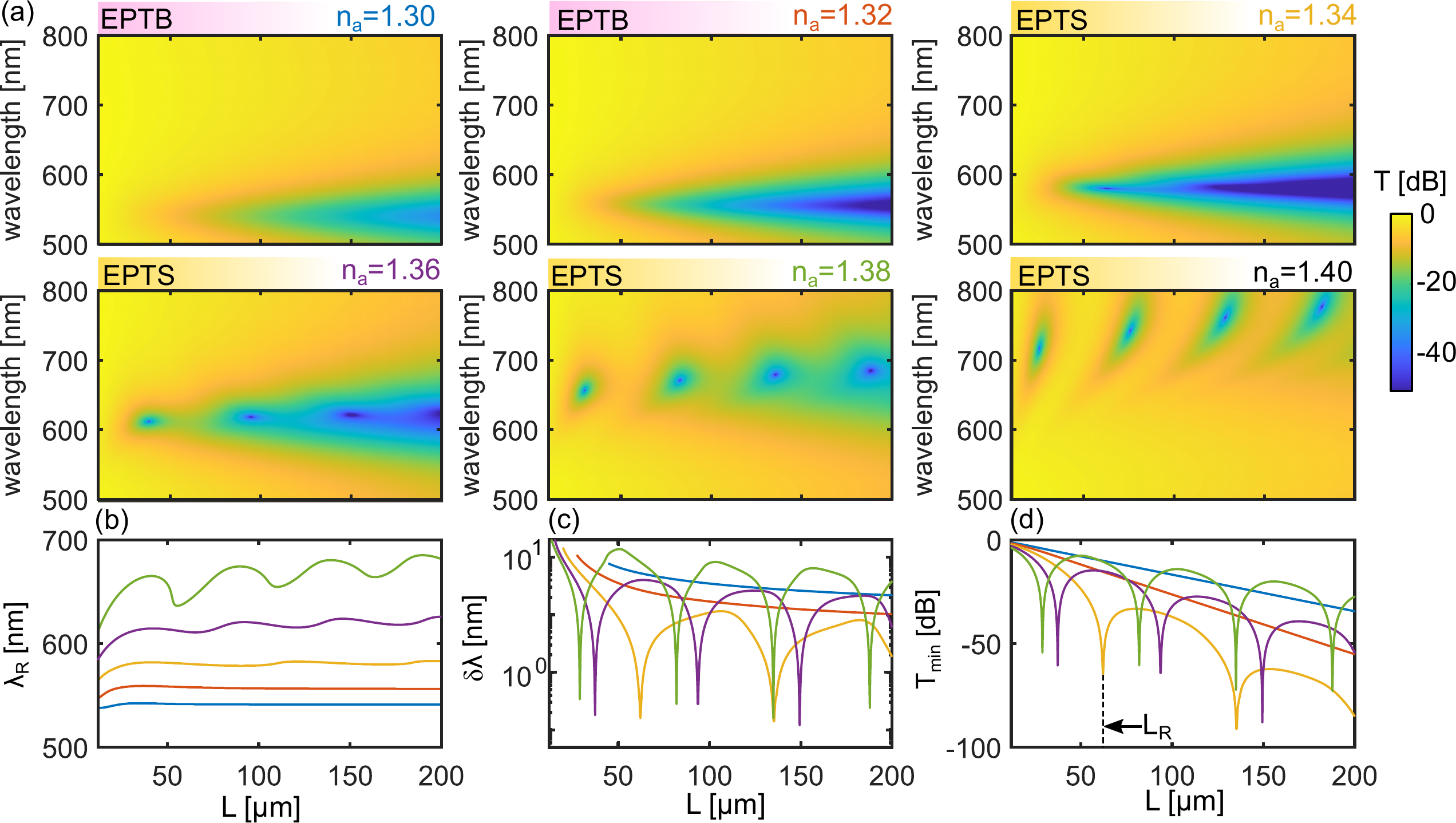}
\caption{(a) Calculated transmission spectra (single mode output) as a function of $L$ and $\lambda$ for different values of $n_a$ as labelled, and (b) associated $\lambda_R$, (c) $\delta\lambda$, (d) $T_{\rm min}$.  In (b)--(d) each curve's colour refers to the $n_a$ labels in (a). For each $n_a$, we identify the first resonance dip at $L_{R}$ (dashed line in (d)), corresponding to half a beat length~\cite{tuniz2016broadband}.}
\label{fig:fig6}
\end{figure*}

\subsection*{Dependence on analyte: relation to sensitivity}

We begin by considering the performance of the sensor as a function of $n_a$ for the representative lengths $L$ considered so far. Henceforth, we will only consider EPTS regime, which can yield the smallest $\delta\lambda$ due to directional coupling. We repeat the calculations for the three analyte lengths shown in Fig.~\ref{fig:fig3}, but with a much finer resolution on $n_a$. The resulting transmission spectra, as a function of $n_a$ and $\lambda$, are shown in Fig.~\ref{fig:fig4}(a)--(c) for $L=25\,\mu{\rm m}$, $L=37.5\,\mu{\rm m}$ , and $L=50\,\mu{\rm m}$ respectively. We immediately note that, although in all cases the resonances qualitatively redshift for increasing analyte index, their shape, sharpness, and location can change significantly. To quantify this further, Fig.~\ref{fig:fig4}(d), (e) and (f) show the associated resonant wavelength $\lambda_R$, 3-dB width $\delta\lambda$, and transmission minimum $T_{min}$ as function of $n_a$, for the same lengths as in (a)--(c):  $L=25\,\mu{\rm m}$ (blue), $L=37.5\,\mu{\rm m}$ (orange), and $L=50\,\mu{\rm m}$ (yellow). Figure~\ref{fig:fig4}(b) shows that the resonant wavelength is dependent on the length of the chosen device, particularly at longer lengths, as anticipated in our  preliminary analysis of Fig.~\ref{fig:fig3}(a) at $n_a =1.38$ and $n_a =1.4$. Furthermore, high analyte indices show a fluctuation in the resonant transmission wavelength as the length of the sensor increases, and even multiple transmission minima -- see also the spectra in Fig.~\ref{fig:fig3}(a), $n_a=1.40$. This behaviour is challenging to interpret, as a result of the wavelength dependence of the complex dispersion profiles, where mode excitation, propagation, and losses all contribute to the total transmission spectrum in  a non-trivial manner. The EPTB regime is dominated by the excitation of the lowest-loss hybrid eigenmode over the entire wavelength range (\ie,  $|a_1|^2 >0.9$), with minimal contributions from the highest-loss mode (\ie, $|a_2|^2 <0.1)$. In contrast, the EPTS regime is characterized by broad wavelength regions (bandwidth: 50-100\,nm) where both modes are excited (i.e., $|a_{1,2}|^2 >0.4$), analogously to earlier reports of broadband plasmonic directional couplers~\cite{tuniz2016broadband}. As a result, fixing the length as per Fig.~\ref{fig:fig4} shows complete coupling only for certain combinations of $n_a$ and $\lambda$; elsewhere, incomplete coupling occurs, and local transmission minima occur over the associated wavelength region, with much larger $\delta\lambda$ values than the optimal case, and fluctuating resonant wavelength $\lambda_R$, as shown in Fig.~\ref{fig:fig4}(c) and the yellow line Fig.~\ref{fig:fig4}(d).

To quantify this effect, Figure~\ref{fig:fig4}(e) shows $\delta\lambda$ as a function of $n_a$:  in the EPTS regime considered, it is possible to achieve a sub-nm $\delta\lambda$ for specific combinations of $L$ and $n_a$. This spectral width minimum is associated with a transmission minimum $T_{min}$, shown in Fig.~\ref{fig:fig4}(f), and would require a high enough SNR to be measured (here: $>60\,$dB).  However, small deviations in $n_a$ can change $\delta\lambda$ significantly for fixed $L$. 

Both the sensitivity $S$ and resonant width $\delta\lambda$ thus depend on device length $L$ and analyte index $n_a$. Therefore, an optimized plasmonic waveguide senor design (small $\delta n$) crucially requires a judicious combination of $n_a$ (chosen by the user) and $L$ (determined by the choice of target $n_a$ range) to ensure that $\delta\lambda$ is minimized. Most importantly, engineering and analysing such designs requires implementing the model presented here to appropriately account for hybrid mode excitation, propagation, and loss.

Recall that while the evaluation of $\delta\lambda$ demands full transmission spectra, recommended design criteria often infer the sensitivity $S$ from modal calculations, as a valuable proxy with fast calculation times~\cite{hassani2007design}, informing the first steps of a plasmonic sensor design. This approach relies on calculating the wavelength $\lambda_i$ at which a particular condition is met, as a function of $n_a$, and computing the associated sensitivity via $S_i = d\lambda_i/dn_a$. Because our fast numerical tool allows us to obtain $\lambda_R$ from $T(\lambda)$, we can now also compare how the sensitivity inferred from different conditions, obtained from mode calculations alone, compares with the $S=d\lambda_R/dn_a$, obtained from propagation calculations. Continuing our earlier analysis, we only consider the EPTS region ($n_a>1.33$), which can achieve the lowest $\delta n$ by complementing high sensitivities with sub-nm $\delta\lambda$ at specific $L$.

Figure~\ref{fig:fig5}(a) shows the loss matching wavelength $\lambda_{LM}$ (purple), the phase-matching wavelength $\lambda_{PM}$ (light blue), the wavelength where real part of the effective index difference is minimum $\lambda_{\Delta n_{\rm eff}^{\rm min}}$ (green), and the plasmonic cutoff wavelength $\lambda_{\rm cutoff}$ (black), as defined in Fig.~\ref{fig:fig2} and  associated text. Note in particular that $\lambda_{PM}$ does not exist for $n_a>1.35$ (\ie, the real parts of the isolated eigenmodes do not cross). Analogously, the imaginary parts of the hybrid eigenmodes do not cross for $n_a>1.385$, and $\lambda_{LM}$ does not exist in that region.  Because $\lambda_R$ depends on $L$, we show $\lambda_R$ for both $L = 25\,\mu{\rm m}$ (dark blue), $L = 37.5\,\mu{\rm m}$ (orange), obtained from the transmission calculations of Fig.~\ref{fig:fig4}. All curves show super-linear wavelength shifts with increasing $n_a$.  The associated sensitivities are shown in Fig.~\ref{fig:fig5}(b) on a logarithmic scale. For the two example lengths shown, we find that the cutoff wavelength of the uncoupled plasmonic mode (black line) appears to be the best proxy for the sensitivity of the full (coupled) dielectric-plasmonic waveguide sensor (blue and orange lines),  predicting the devices' sensitivity to within a factor of $\sim 3$ over the entire $n_a$ range. However, this is not the case if a  wider range of lengths $L$ is considered: The shaded grey region in Fig.~\ref{fig:fig5}(a) and Fig.~\ref{fig:fig5}(a) respectively show the ranges of possible $\lambda_R$ and associated sensitivities- for $L=10-2000\,\mu {\rm m}$ -- the latter of which spans more than an order of magnitude. This example serves to reinforce the message that no single criterion obtained from mode dispersion calculations can be used to accurately infer device sensitivity, which is crucially dependent on the device length.

\subsection*{Dependence on length: relation to the exceptional point}

We now calculate the transmission spectra for  representative values of $n_a$, but with a much finer resolution on $L$. The resulting transmission spectra, as a function of $L$ and $\lambda$, are shown in Fig.~\ref{fig:fig6}(a). Figure~\ref{fig:fig6}(b), (c) and (d) respectively show the associated resonant wavelength $\lambda_R$, 3-dB width $\delta\lambda$, and transmission minimum $T_{min}$  as function of $L$ for each analyte index $n_a$ of Fig.~\ref{fig:fig6}(a) as labelled. Figure~\ref{fig:fig6}(b) confirms that the resonant wavelength is nominally constant for all $L$ and $n_a\leq 1.33$ (in the EPTB regime). For $n_a>1.33$ (in the EPTS regime), however, $\lambda_R$ is dependent on length and can fluctuate significantly, as per our earlier analysis.

Figure~\ref{fig:fig6}(c) shows that the spectral width $\delta\lambda$ decreases monotonically with increasing length for resonances occurring in the EPTB regime (\ie, $n_a = 1.30$ and $n_a = 1.32$), because only coupling to the lowest-loss mode dominates, and resonant interference does not occur. In contrast, the EPTS regime supports resonant coupling effects from the dielectric core to the gold surface -- or equivalently, resonant interference between the two hybrid modes --  leading to sub-nm $\delta\lambda$ at specific interaction lengths $L$. It is striking to note that in the EPTS region, the wavelength at which complete coupling occurs (\ie, blue colourmap regions in Fig.~\ref{fig:fig6}(a)) can also depend on $L$. This effect becomes more prominent as $n_a$ increases, see for example the case of $n_a=1.4$, and is due to the differential wavelength-dependent losses between the two beating modes, leading to a change in the wavelength with most complete destructive interference.
Once again, $\delta\lambda$ minima are associated with local $T_{min}$ minima as shown in Fig.~\ref{fig:fig6}(c). In the EPTB regime a monotonic decrease in $T_{min}$ is observed, consistent with a behaviour dominated by the lowest loss mode alone. In the EPTS regime however $T_{min}$ oscillates as a result of directional coupling. 

\begin{figure*}[t!]
\centering
\includegraphics[width=\textwidth]{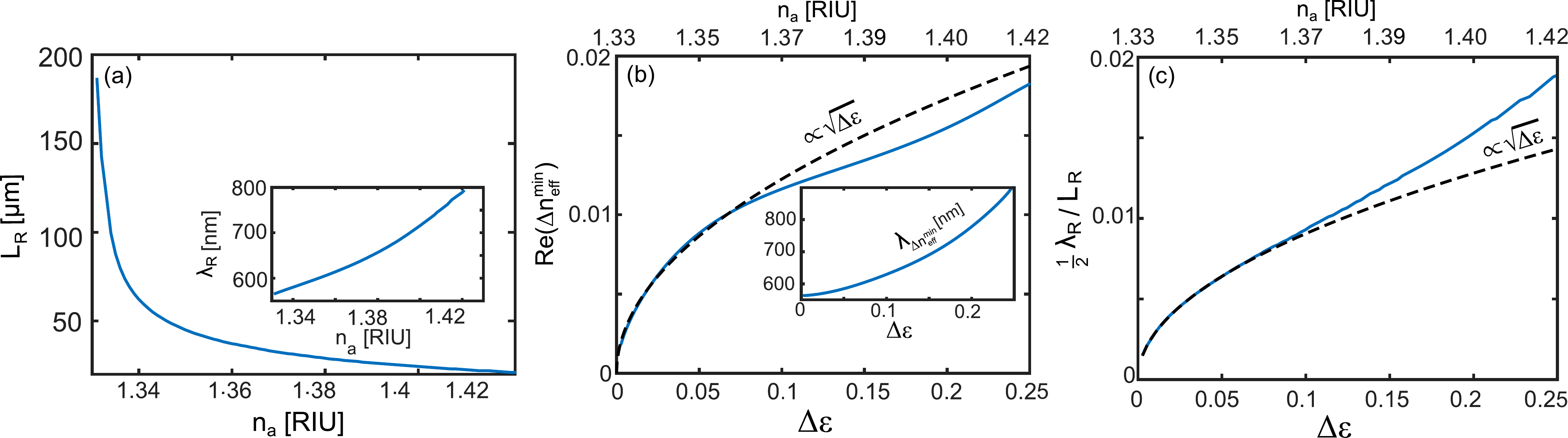}
\caption{(a) First resonance dip $L_{R}$ obtained from Fig.~\ref{fig:fig6}(d). The wavelength $\lambda_R$ at which it occurs is shown in the inset. Note the rapid increase in $L_R$ close $n_a=1.33$, corresponding to the EP.  (b) Eigenmode splitting as a function of analyte permittivity perturbation $\Delta\varepsilon$ with respect to the EP at $n_a=1.33$, obtained from the dispersion of Fig.~\ref{fig:fig2}, and showing a characteristic square-root dependence. Inset: wavelength dependence of the eigenmode splitting. (c) Eigenmode splitting estimated from the calculated transmission spectra of Fig.~\ref{fig:fig6}. The ratio $\frac{1}{2}\lambda_R/L_{R}$ follows a square root dependence on $\Delta\varepsilon$ for small perturbations. }
\label{fig:fig7}
\end{figure*}

Let us now analyze how directional coupling is affected by $n_a$, and its relation to the exceptional point. To simplify the discussion, we consider $L_{R}$ as the location of the first transmission minimum, highlighted by a dotted line in Fig.~\ref{fig:fig6}(d) for the case $n_a=1.34$, which roughly (but not exactly) corresponds to half a beat length (the two differ because of modal losses). Inspecting Fig.~\ref{fig:fig6}(d) already suggests that $L_{R}$ depends on $n_a$, and Figure~\ref{fig:fig7}(a) quantitatively plots this dependence. As per our earlier analysis, the inset of Fig.~\ref{fig:fig7}(a) shows that the wavelength at which $L_{R}$ occurs is itself a function of $n_a$. Figure~\ref{fig:fig7}(a) clearly shows that $L_{R}$ increases dramatically close to the exceptional point at $n_a=1.33$, and gradually reduces as the analyte index increases. We now show that this property is a direct consequence of the dispersion topology of Fig.~\ref{fig:fig2}(e), and indeed manifests key properties of perturbations near the exceptional point.

Sensors which operate close to the exceptional point have attracted much attention in recent years,  because the splitting of the coalesced eigenvalues is proportional to the \emph{square root} of the change in the environment's relative permittivity $\Delta \varepsilon$, with a much higher slope than the \emph{linear} splitting for conventional Hermitian systems~\cite{wiersig2020prospects}, and can thus be exploited for enhanced sensing~\cite{chen2017exceptional, zhang2019noninvasive, wiersig2020prospects}. The difference in effective index can be measured using an interferometric setup as a resonant dip with wavelength exquisitely dependent on analyte index  - which one would expect is exactly what happens in our sensor through the beating between the two modes over a finite length. However, near the exceptional point the small splitting corresponds to beat lengths much larger than the loss length of the modes, so that dips are dominated by loss, rather than interference, and the benefits of the EP are lost. Intriguingly, to restore the benefits one could consider  compensating losses by adding gain -- which would get the sensor geometry closer to a true \PT-symmetric system.

Nonetheless, the characteristic square root dependence near the EP  is readily found from the hybrid eigenmode calculations of Fig.~\ref{fig:fig2}. To illustrate this, Figure~\ref{fig:fig7}(b) shows a plot of the associated minimum eigenvalue splitting, \ie, $\Re e(\Delta n_{\rm eff}^{\rm min}) = min[\Re e( n_{\rm eff,2} - n_{\rm eff,1})]$, for each $\Delta \varepsilon = n_a^2-n_{EP}^2$. For small values of $\Delta\varepsilon$, the eigenvalue splitting follows a square root dependence (black dashed line). For large values of $\Delta\varepsilon$, this is no longer the case because perturbations to the EP are large. Note that $\Re e(\Delta n_{\rm eff}^{\min})$ occurs at wavelengths longer than the cutoff of the isolated plasmonic mode -- a further indication of how removed this geometry becomes from the  perturbative treatment.

One important subtlety, in the present context, is that the wavelength at which the minimum eigenmode splitting occurs, shown in the inset of Fig.~\ref{fig:fig7}(b), is also a function of  $\Delta\varepsilon$. Recalling that the beat length and eigenmode splitting are related by $L_b = \lambda_R/\Delta n_{\rm eff}$ -- at least in the lossless case -- Fig.~\ref{fig:fig7}(c) shows $\frac{1}{2}\lambda_R/L_{R}$ as a function of $\Delta\varepsilon$, as obtained from the transmission spectra of Fig.~\ref{fig:fig6}. We find a square root dependence (black dashed line) for small $\Delta\varepsilon$ perturbations, and deviations for larger values of $n_a$: the signature square-root dependence on perturbation, close to the EP, is thus obtained from the device beat length.  Interestingly, the deviation from the square root dependence is opposite to that predicted from mode calculations alone at high index region, once again as a result of the subtle interplay of mode excitation, interference, and loss -- a further indication that mode dispersion calculations, in isolation, do not adequately predict the behaviour of plasmonic sensors.

\section{APPLICABILITY TO SHORT-RANGE PLASMONS}
\label{sec:SRSPP}

\begin{figure*}[t!]
\centering
\includegraphics[width=0.9\textwidth]{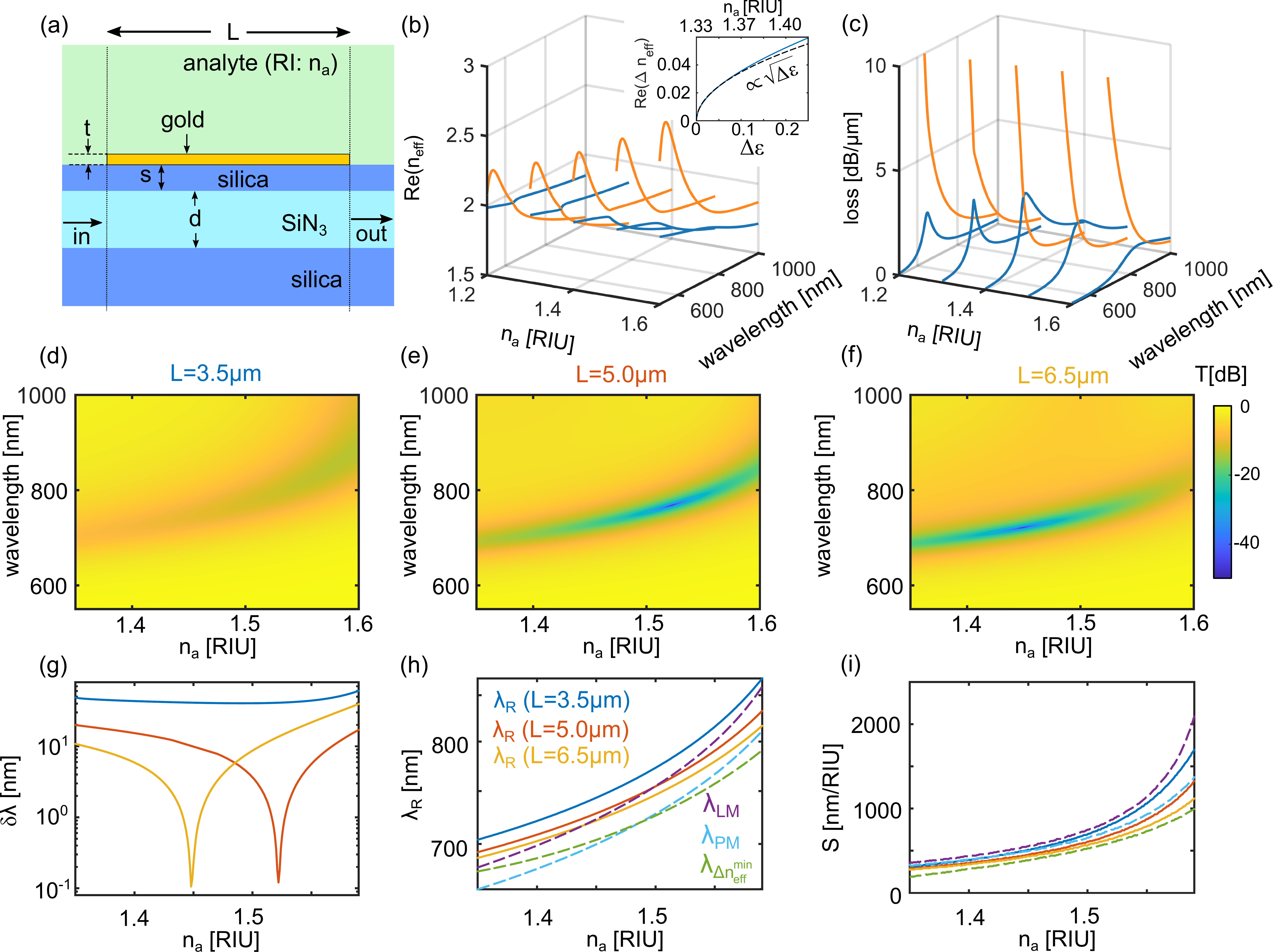}
\caption{(a)  Concept schematic of a chip-scale waveguide sensor: a silicon nitride (SiN$_3$) core ($d=400\,{\rm nm}$) is phase matched to a short-range plasmon plasmon mode of a gold film ($t=30\,{\rm nm}$) in the visible. Here we consider a silica spacer of $s=111\,{\rm nm}$ leads to an EP for $n_a=1.33$ at a wavelength of 670\,nm. Also shown are a detailed 3D plot of the associated (a) $\Re e(n_{\rm eff}$) and (b) loss a function  of wavelength and $n_a$, showing a transition from the EPTB to EPTS regime by increasing the $n_a$. Inset in (b): square root dependence of $Re (\Delta n_{\rm eff})$ on the change in analyte permittivity with respect to the EP at $n_a=1.33$. Also shown are the calculated transmission spectra (single mode output) as a function of $n_a$ and $\lambda$ for different values of (c) $L=2.5\,\mu{\rm m}$ (blue), (d) $L=5.0\,\mu{\rm m}$ (orange) and (e) $L=7.5,\mu{\rm m}$ (yellow). (f) Associated (g) $\delta\lambda$, (h)  $\lambda_R$, and (d) $S$. Line colours in (g)--(i) correspond to that of the $L$ values labelled in (d)--(f). For comparison, the characteristic wavelengths obtained from mode dispersion calculations, and the resulting $S$, are shown as dashed lines in (h) and (i) respectively, where $\lambda_{LM}$: purple; $\lambda_{PM}$: light blue; $\lambda_{\Delta{n_{eff}^{min}}}$: green.}
\label{fig:fig8}
\end{figure*}

Having considered  a plasmonic sensor which relies on the long-range surface plasmon, it is worth briefly considering how our conclusions carry over to the case of  short-range surface plasmons~\cite{fan2012refractive, fan2013integrated}, which have no cutoff, and are characterized by larger effective index, lower effective modal area, and lower group velocities, providing a pathway for nanometer-scale optical confinement on chip-compatible geometries, at the cost of higher losses. In the present context, phase matching to the SR-SPP thus requires different geometry than in previous sections, namely a higher-index core waveguide adjacent to the plasmonic film. We consider the structure shown in the schematic of Figure~\ref{fig:fig8}(a), composed of a dielectric waveguide formed by a silicon nitride core (core width: $d=400\,{\rm nm}$; refractive index dispersion: Ref.~\cite{luke2015broadband}), surrounded by an infinite silica layer on one side, and a gold nanofilm of length $L$ on the other ($t=30\,{\rm nm}$), separated by a silica spacer. As per our analysis so far, we wish to detect changes to the refractive index $n_a$ of an analyte surrounding the waveguide. We choose a spacer thickness of $s=111\,{\rm nm}$,  such that the the exceptional point is at $n_a = 1.33$ and at $\lambda=670\,{\rm nm}$. The calculated $\Re e(n_{\rm eff})$ and loss, as a function of wavelength and analyte index, are shown in Fig.~\ref{fig:fig8}(b) and (c) respectively. Note that compared to the structure of Fig.~\ref{fig:fig1}(a), whose modal dispersion curves are shown in Fig.~\ref{fig:fig2}(e), this geometry has larger effective index and loss.

Figure~\ref{fig:fig8}(d),(e) and (f) show colourplots of the calculated transmission spectrum as a function of $n_a$ for three example lengths $L=3.5\,\mu{\rm m}$, $5.0\,\mu{\rm m}$ and $7.5\,\mu{\rm m}$ respectively, using the EM method. We can immediately identify that many of the features discussed so far carry over: both the spectral width $\delta\lambda$, shown in Fig.~\ref{fig:fig8}(g), and resonant wavelength $\lambda_R$, shown in Fig.~\ref{fig:fig8}(h), depend on the choice of the interaction length. The resulting sensitivity, shown in Fig.~\ref{fig:fig8}(i), also depends on $L$, and can vary by nearly a factor of 2, especially in regions of higher sensitivity and at higher analyte indices. Note in particular that the region of highest sensitivity occurs for the shortest length $L=3.5\,\mu{\rm m}$ (Fig.~\ref{fig:fig8}(i) blue line), which however has a $\delta\lambda$ that is nominally one order of magnitude larger than that for  the other two lengths, thus leading to lower detection limits. Furthermore, regions of small $\delta\lambda$ occur for different combinations of $n_a$ and $L$, since they are linked to the characteristic beat length for a certain configuration. Finally, we find that $\lambda_{PM}$, $\lambda_{LM}$, and $\lambda_{\Delta{n_{eff}^{min}}}$ (dashed curves in Fig.~\ref{fig:fig8}(h)), obtained from the dispersion curves of Fig.~\ref{fig:fig8}(b),(c), do not generally correspond to the resonant wavelengths obtained from Fig.~\ref{fig:fig8}(d)--(e). 
In all cases, the associated sensitivities (Fig.~\ref{fig:fig8}, dashed curves) also vary significantly, meaning that an \emph{a-priori} quantitative analysis on the basis of mode dispersion calculations provide an order-of-magnitude estimate only, and that device performance requires a case-by-case analysis that considers propagation through the sensor. Recent chip-based waveguide sensors harnessing plasmonic Mach-Zehnder interferometry have shown that the resonance due to directional coupling can be fine-tuned by appropriately phase-shifting a reference waveguide arm~\cite{chatzianagnostou2019scaling} -- which in the present context could be adapted to target a minimum $\delta\lambda$ in regions of highest $S$.

\section{Conclusion}
In conclusion, we have comprehensively evaluated the sensing properties of plasmonic waveguide sensors by calculating their resonant transmission spectra in different regions of the non-Hermitian eigenmode space. Our study highlights the limits of using modal dispersion calculations alone to predict plasmonic sensor performance and transmission spectra. These limits are easily addressed by using the same modal calculations in the framework of a model which accounts for both excitation and propagation of the eigenmodes supported by the sensor. The resulting transmission calculations faithfully reproduce the transmission spectra, verified via a comparison with full-vector finite element calculations, with the added benefit of allowing for a rapid sweep over three important parameters (wavelength and analyte index, but most importantly device length), in turn revealing many important aspects that have so far eluded discussion in the context of practical devices. By increasing the resolution on the full wavelength- and analyte- parameter space, we showed that no single mode dispersion criterion can be used as a proxy for sensitivity. Indeed, the highest detection limits occur where directional coupling is supported (via sub-nm spectral linewidths) and close to plasmonic cutoffs. The latter suggests revisiting sensor performance in cylindrical fibers/wires, close to the cutoff of high-order long-range cylindrical plasmonic modes. Near the exceptional point the hybrid plasmonic modes modes yield a characteristic square root dependence of the eigenmode splitting with respect to the permittivity perturbation of the sensor, which in this context is  identified through the sensor beat length.
The square root dependence theoretically leads to high slopes of the coupling length {\em vs} refractive index, and thus high sensitivity. However, the small difference in effective index between beating modes in this supralinear region near the exceptional point corresponds to  beat lengths that are much larger than the modes' loss length. In this case, the transmission dip is dominated by loss rather than interference. This makes it difficult to exploit the supralinear behaviour near the exceptional point in lossy systems. Theoretically, this issue could be avoided if one could compensate losses by introducing gain, thereby bringing the sensor closer to a true \PT-symmetric system. 
Note that recent experiments on a nanofluidic-core fiber platforms~\cite{gomes2021direct} showed that hybrid mode excitation and propagation can be directly visualized via sideways-detected fluorescence, which could be adapted in the present context for novel, single-wavelength sensing avenues that rely on beat-length measurements. Our analysis will find widespread applications in a variety of waveguide-based refractive index sensors, whose theoretical performance, in some instances, might require revisiting.

\noindent\textbf{Funding.} Australian Research Council Discovery Early Career Researcher Award (DE200101041).

\noindent\textbf{Acknowledgements.} A.T. and B.T.K. thank Pranav A. Alavandi and Zachary J. R. Mann for fruitful discussions.

\noindent\textbf{Disclosures.} The authors declare no conflicts of interest.

\noindent\textbf{Data Availability.} Data underlying the results presented in this paper are not publicly available at this time but may be obtained from the authors upon reasonable request.

\bibliography{sample}

\bibliographyfullrefs{sample}


\end{document}